\documentclass{elsart4}

\usepackage{amssymb}

\usepackage{graphics}
\usepackage{graphicx}
\usepackage{epsfig}
\usepackage{amssymb}
\journal{Physica B}

\begin{document}

\begin{frontmatter}

 \title{Rate of equilibration of a one-dimensional Wigner crystal}
 \author[argonne]{K. A. Matveev\corauthref{cor1}},
 \author[seattle]{A. V. Andreev},
 \author[georgia]{M. Pustilnik}
 \address[argonne]{Materials Science Division, Argonne National Laboratory,
            Argonne, Illinois 60439, USA}
 \address[seattle]{Department of Physics, University of Washington, Seattle, 
            Washington 98195, USA}
\address[georgia]{School of Physics, Georgia Institute of Technology, Atlanta,
            Georgia 30332, USA}
 \corauth[cor1]{Corresponding author. E-mail: matveev@mailaps.org}

\begin{abstract}
  We consider a system of one-dimensional spinless particles interacting via
  long-range repulsion.  In the limit of strong interactions the system is a
  Wigner crystal, with excitations analogous to phonons in solids.  In a
  harmonic crystal the phonons do not interact, and the system never reaches
  thermal equilibrium.  We account for the anharmonism of the Wigner crystal
  and find the rate at which it approaches equilibrium.  The full
  equilibration of the system requires umklapp scattering of phonons,
  resulting in exponential suppression of the equilibration rate at low
  temperatures.
\end{abstract}

\end{frontmatter}

The low-temperature physics of interacting electron systems is usually
described in the framework of the so-called Luttinger liquid theory
\cite{haldane}.  The phenomenological nature of this approach enables one to
study the systems with any interaction strength, provided that the physics is
controlled by the low-energy excitations.  On the other hand, even at low
temperature $T$ some phenomena involve excitations with energies much higher
than $T$.  In such cases microscopic approaches are usually more effective.

\begin{figure}
\includegraphics[width=.47\textwidth]{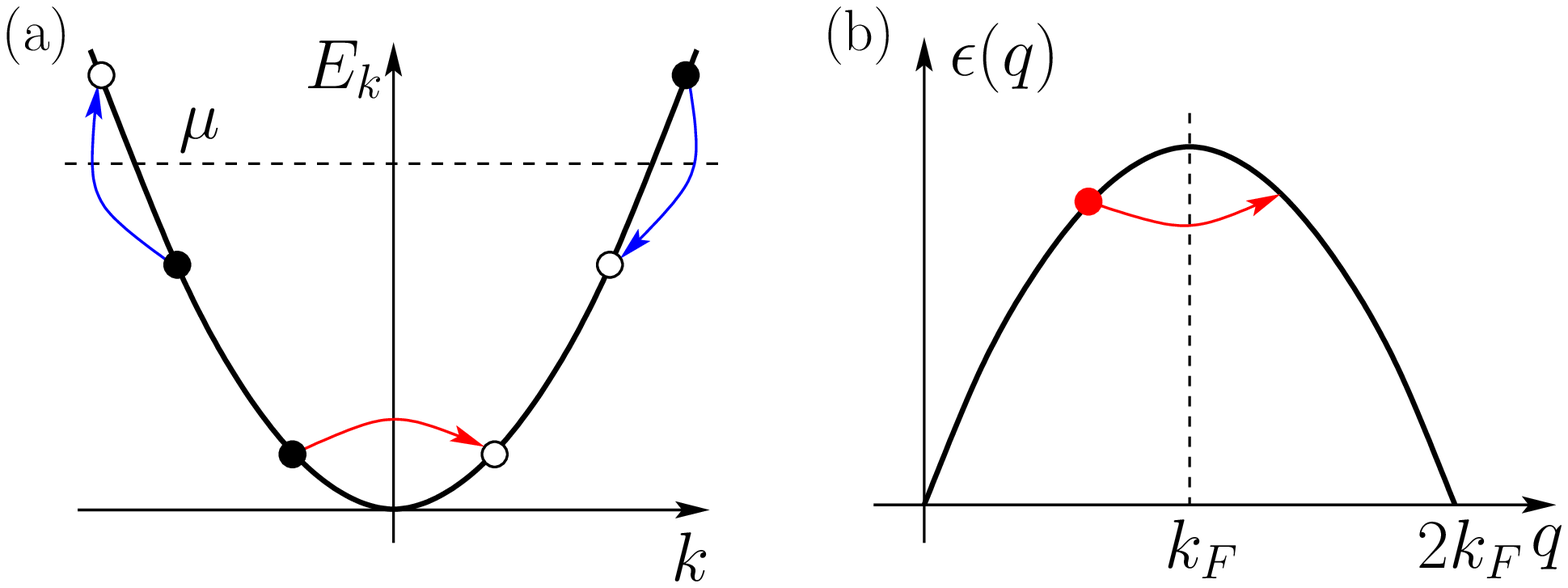}
\caption{(a) Three particle scattering process leading to equilibration of
  weakly-interacting one-dimensional electrons.  (b) Spectrum of the hole
  excitation.  Backscattering of the hole occurs when its wave vector crosses
  the point $q=k_F$.}
\label{fig:three-particle}
\end{figure}

An example of such a phenomenon is the equilibration of a one-dimensional
system.  The latter is understood most easily in the case of weak interactions
between electrons, when the usual picture of quasiparticle excitations is
applicable.  Because of the conservation of momentum and energy in
electron-electron scattering, the two-particle collisions do not affect the
distribution function.  Thus at weak interactions the equilibration is
accomplished via the three-particle collisions,
Fig.~\ref{fig:three-particle}(a).  To equilibrate the chemical potentials of
the right- and left-moving electrons, the collisions must change the numbers
of electrons on each branch, i.e., the backscattering events, such as the one
shown in Fig.~\ref{fig:three-particle}(a), are needed.  The most effective
such process includes backscattering of a particle at the very bottom of the
band \cite{lunde}.  The small probability of finding an available final state
deep below the Fermi level $E_F$ results in exponential suppression of the
equilibration rate $\tau^{-1}\propto e^{-E_F/T}$ \cite{micklitz}.

The key event in the process of equilibration is the backscattering of a hole
at the bottom of the band.  It is therefore helpful to focus on the motion of
such a hole in momentum space.  For electrons with quadratic spectrum
$E_k=\hbar^2k^2/2m$ the energy of the hole excitation is
$\epsilon(q)=E_{k_F}-E_{k_F-q}=\hbar v_F q(1-q/2k_F)$, assuming that the hole
is created by moving an electron from state $k_F-q$ to the right Fermi point
$k_F$.  Here $k_F$ is the Fermi wave vector and $v_F=\hbar k_F/m$ is the Fermi
velocity.  Backscattering occurs when the wave vector of the hole $q$ crosses
$k_F$, Fig.~\ref{fig:three-particle}(b).

Because the equilibration of the system involves excitations with energies of
the order of the bandwidth, this phenomenon is not captured by the
conventional Luttinger liquid theory \cite{haldane}.  This makes the
generalization of the above picture beyond the weakly-interacting limit rather
challenging.  On the other hand, it is possible to develop a microscopic
theory of this phenomenon in the case of strong Coulomb interactions
\cite{equilibrationWigner}.  Here we revisit this approach and obtain the full
expression for the equilibration rate for arbitrary strong long-range
repulsion.  The key idea is that as repulsion of electrons becomes stronger,
the system minimizes its energy by forming a periodic structure known as the
Wigner crystal.  Although the long-range order in such a system is destroyed
by quantum fluctuations \cite{schulz}, the presence of the strong short-range
order enables us to treat the system as anharmonic chain described by the
Hamiltonian
\begin{equation}
  \label{eq:H}
  H=\sum_l \frac{p_l^2}{2m} + \frac12\sum_{l,l'} V(x_l-x_{l'}).
\end{equation}
Here $p_l$ and $x_l$ are the momentum and coordinate of the $l$-th particle
and $V(x)$ is the interaction potential.

\begin{figure}
\includegraphics[width=.47\textwidth]{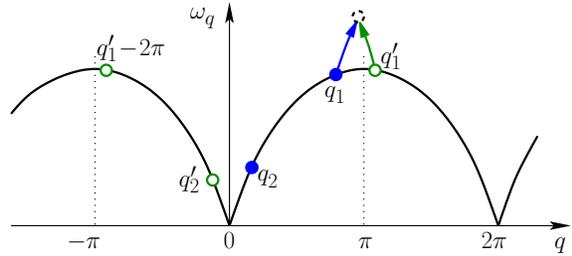}
\caption{Excitation spectrum of a one-dimensional Wigner crystal.
  Equilibration processes involve a phonon $q_1$ being scattered to a state
  $q_1'$ outside the Brillouin zone as a result of a collision with a thermal
  phonon $q_2$.  Conservation of energy and momentum requires the latter to be
  backscattered to $q_2'$. This umklapp process is analogous to a hole
  crossing the point $q=k_F$ in the weakly interacting Fermi gas,
  Fig.~\ref{fig:three-particle}(b).  (Note that if the distances are measured
  in units of $a$, the Fermi wave vector $k_F=\pi$.)}
\label{fig:spectrum}
\end{figure}

The excitations of the system are essentially the phonons in the electronic
crystal.  They are conveniently described in terms of the displacements
$u_l=x_l-la$ of electrons from their equilibrium positions, where $a$ is the
mean interparticle distance.  Strong repulsion means small displacements,
$|u_l-u_{l'}|\ll |l-l'|a$, and the Hamiltonian can be approximated by that of
a harmonic chain
\begin{equation}
  \label{eq:H_0}
  H_0=\sum_l \frac{p_l^2}{2m} 
      + \frac14\sum_{l,l'} V_{l-l'}^{(2)}\,(u_l-u_{l'})^2,
\end{equation}
where the $r$-th derivative of $V(x)$ is denoted as
\begin{equation}
  \label{eq:V^(n)}
  V_l^{(r)}=\left.\frac{d^r V(x)}{dx^r}\right|_{x=la}.
\end{equation}
The phonon modes of this Hamiltonian are easily found,
\begin{equation}
  \label{eq:omega_q}
  \omega_q^2=\frac{2}{m}\sum_{l=1}^\infty V^{(2)}_l [1-\cos(ql)].
\end{equation}
At small wave vector $qa\to0$ the excitation spectrum is linear,
$\omega_q=s|q|$, where $s=(\sum_l V_l^{(2)}l^2/m)^{1/2}$ is the ``sound
velocity'' in the Wigner crystal measured in units of lattice spacings per
unit time.  The spectrum is periodic in $q$, with the Brillouin zone
$-\pi<q<\pi$, Fig.~\ref{fig:spectrum}

Any eigenstate of the harmonic chain (\ref{eq:H_0}) can be described by the
set of occupation numbers $N_q$ of the phonons.  In the absence of interaction
of phonons, the lifetime of any such state is infinite, and the system never
reaches thermal equilibrium.  On the other hand, the harmonic Hamiltonian
(\ref{eq:H_0}) is merely the leading term of the expansion of Eq.~(\ref{eq:H})
in the small parameter $K=\pi\hbar/ma^2s$.  The next term is proportional to
the third power of the displacement $u_l$ and generates scattering processes
involving three phonons.  At low temperature $T\ll \hbar \omega_\pi$ the
typical quasimomenta of phonons are small, $q\ll T/\hbar s$, umklapp
scattering is suppressed, and the phonons remain in the first Brillouin zone
upon scattering.  This means that apart from energy, collisions conserve the
total quasimomentum of the phonons.  This yields the equilibrium phonon
distribution
\begin{equation}
  \label{eq:phonon_wind}
  N_q=\frac{1}{e^{\hbar(\omega_q-uq)/T}-1}
\end{equation}
characterized by two parameters, the temperature $T$ and the velocity $u$ of
the phonon gas with respect to the crystal.

On the other hand, even at low temperatures there are rare umklapp collisions
of phonons, such as the one shown in Fig.~\ref{fig:spectrum}, which do not
conserve their quasimomentum.  As a result, one expects the velocity $u$ to
relax gradually as $\dot u=-u/\tau$ with a small relaxation rate $\tau^{-1}$.
To find it we notice that the most efficient umklapp processes involve a
phonon $q_1$ near the boundary $q=\pi$ of the Brillouin zone colliding with a
thermal phonon with $q_2\sim T/\hbar s$, Fig.~\ref{fig:spectrum}.  At low
temperature the resulting change of quasimomentum $|q_1'-q_1|\sim T/\hbar s$
is small compared to the typical scale $q_T\sim (T/\hbar|\omega''_\pi|)^{1/2}$
of the distribution (\ref{eq:phonon_wind}) near the edge of the Brillouin
zone.  (Here $\omega''_q=\partial_q^2\omega_q$.)  Thus the high-energy phonon
performs a slow diffusive motion in the momentum space, and its distribution
function $N_q(t)$ obeys the Fokker-Planck equation
\begin{equation}
  \label{eq:Fokker-Planck}
  \partial_t N_q = \partial_q \left[\frac{B(q)}{2}
                   \left(\frac{\hbar\omega_q'}{T}+\partial_q\right)
                             \right]N_q.
\end{equation}
Here 
\begin{equation}
   \label{eq:B_definition}
   B(q)=\sum_{\delta q} (\delta q)^2W_{q,q+\delta q}
\end{equation}
has the meaning of the diffusion constant in momentum space and $W_{q,q+\delta
  q}$ is the rate at which a phonon $q$ changes its wave vector by $\delta q$
as a result of collisions with other phonons.

The Fokker-Planck equation should be solved with the boundary conditions
\begin{equation}
  \label{eq:boundary_conditions}
  N_q=e^{-\hbar\omega_q/T} e^{\pm\pi\hbar u/T},
\quad
  q_T\ll \mp(q-\pi)\ll\pi
\end{equation}
obtained by extending the distribution (\ref{eq:phonon_wind}) beyond the first
Brillouin zone.  Such solution \cite{equilibrationWigner} gives the relaxation
law $\dot u=-u/\tau$ with the rate
\begin{equation}
  \label{eq:relaxation_rate}
  \tau^{-1}=3B\left(\frac{\hbar s}{T}\right)^3
                 \left(\frac{\hbar|\omega''_\pi|}{2\pi T}\right)^{1/2}
                 e^{-\hbar\omega_\pi/T},
\end{equation}
where $B=B(\pi)$.  The temperature dependence of the relaxation rate is
dominated by the exponentially small probability of the occupation of phonon
states near the edge $q=\pi$ of the Brillouin zone.  Expression
(\ref{eq:relaxation_rate}) is analogous to the result $\tau^{-1}\propto
e^{-E_F/T}$ for weakly-interacting electrons.  The strong interactions between
electrons renormalize the activation temperature from $E_F$ to
$\hbar\omega_\pi$ in Eq.~(\ref{eq:relaxation_rate}).

The temperature dependence of the prefactor in Eq.~(\ref{eq:relaxation_rate})
is determined by that of the diffusion constant $B$ and by $T^{-7/2}$
explicitly present in (\ref{eq:relaxation_rate}).  The former can be deduced
phenomenologically \cite{equilibration_spinless} by treating the phonon near
$q=\pi$ as a mobile impurity in a Luttinger liquid, for which $B$ is known
\cite{castroneto} to scale as
\begin{equation}
  \label{eq:B_vs_T}
  B=\chi T^5,
\quad
  T\to0.
\end{equation}
We therefore conclude that the equilibration rate scales with temperature as
$\tau^{-1}\propto T^{3/2}e^{-\hbar\omega_\pi/T}$.

The constant $\chi$ in Eq.~(\ref{eq:B_vs_T}) has to be determined by
microscopic evaluation of the scattering rate $W_{q,q+\delta q}$ in
Eq.~(\ref{eq:B_definition}).  The dominant scattering process, illustrated in
Fig.~\ref{fig:spectrum}, involves two phonons in both the initial and final
states.  Such scattering can be accomplished either in the first order in
four-phonon scattering amplitude or in the second order in three-phonon
scattering amplitude.  The resulting expression for the scattering rate has
the form \cite{equilibrationWigner}
\begin{eqnarray}
  \label{eq:W}
  W_{q_1^{},q_1'}&=&\frac{2\pi\hbar^2}{m^6N^2}
               \sum_{q_2,q_2'} \frac{\Lambda^2 N_{q_2}(N_{q_2'}+1)}
                   {\omega_{q_1}\omega_{q_2}\omega_{q_1'}\omega_{q_2'}}\,
                 \delta_{q_1+q_2,q_1'+q_2'}
\nonumber\\
              &&\times
                 \delta(\omega_{q_1}+\omega_{q_2}
                      -\omega_{q_1'}-\omega_{q_2'}).
\end{eqnarray}
Here $N$ is the total number of particles in the system and
\begin{eqnarray}
  \label{eq:Lambda}
  \Lambda&=&
                 -\frac{f_3(q_1,q_2)f_3(q_1',q_2')}
                       {\omega^2_{q_1+q_2}-(\omega_{q_1}+\omega_{q_2})^2}
+\frac{f_3(q_2,-q_1')f_3(q_1,-q_2')}
                       {\omega^2_{q_2-q_1'}-(\omega_{q_2}-\omega_{q_1'})^2}
\nonumber\\
               &&+\frac{f_3(q_1,-q_1')f_3(q_2,-q_2')}
                       {\omega^2_{q_2-q_2'}-(\omega_{q_2}-\omega_{q_2'})^2}
+\frac{m}{2}f_4(q_1,q_2,-q_1'),
\end{eqnarray}
with the functions $f_3$ and $f_4$ defined as

\begin{eqnarray*}
%  \label{eq:f_3}
  f_3(q_1,q_2)&=& \sum_{l=1}^\infty V^{(3)}_l 
                 \{\sin[(q_1+q_2)l]
\\
                 &&-\sin(q_1l) -\sin(q_2l)\},
\\
  f_4(q_1,q_2,q_3)&=& \sum_{l=1}^\infty V^{(4)}_l 
                     \{1-\cos(q_1l)-\cos(q_2l)
\nonumber\\
                  &&-\cos(q_3l)-\cos[(q_1+q_2+q_3)l]
\nonumber\\
                  &&+\cos[(q_1+q_2)l]+\cos[(q_1+q_3)l]
\nonumber\\
                  &&+\cos[(q_2+q_3)l]\}.
\end{eqnarray*}

The expressions (\ref{eq:W}) and (\ref{eq:Lambda}) are valid for any $q_2$ and
$q_2'$.  At low temperature their magnitudes are small, $|q_2|, |q_2'|\lesssim
T/\hbar s$.  Thus to find $\chi$ in Eq.~(\ref{eq:B_vs_T}) we need to expand
$\Lambda$ in powers of $q_2$ and $q_2'$.  Carrying out such expansion and
taking into account conservation of momentum and energy we find
\begin{eqnarray}
  \label{eq:Lambda_expanded}
  \Lambda&=&(\delta q)^2\frac{m^2n^4\omega_{q}}{8v^2}\Upsilon_q,
\\
  \label{eq:Upsilon}
  \Upsilon_q&=&(\partial_n\omega_q)\partial_n(v^2-v_q^2)
              -(v^2-v_q^2)\partial_n^2\omega_q
\nonumber\\
          &&
            +\omega_q'' (\partial_n\omega_q)^2n^{-2},
\end{eqnarray}
where $\delta q=q_2-q_2'$ is the small momentum change as a result of
scattering, $n=1/a$ is the particle density, $v=s/n$ and $v_q=\omega_q'/n$ are
the physical velocities of the phonons with wave vectors $0$ and $q$,
respectively.  To leading order in $\delta q$ we have replaced $q_1,q_1'\to
q$.

Equations (\ref{eq:W}), (\ref{eq:Lambda_expanded}), and (\ref{eq:Upsilon})
determine the scattering rate $W_{q,q+\delta q}$ for any $q$.  It is easy to
see that $W_{q,q+\delta q}\propto (\delta q)^2$, and
Eq.~(\ref{eq:B_definition}) immediately gives the temperature dependence
(\ref{eq:B_vs_T}).  To find the coefficient $\chi$ we set $q=\pi$ and obtain
% We now substitute (\ref{eq:Lambda_expanded}) into (\ref{eq:W}) to find
% $W_{q,q+\delta q}\propto (\delta q)^2$.  Then Eq.~(\ref{eq:B_definition})
% immediately gives the temperature dependence (\ref{eq:B_vs_T}).  The
% coefficient $\chi$ is found for any $q$,
% \begin{equation}
%   \label{eq:chi(q)}
%   \chi(q)=\frac{F(v_q/v)\Upsilon_q^2}{2\pi\hbar^3m^2v^{10}(v^2-v_q^2)},
%  \end{equation}
% where
% \begin{equation}
%   \label{eq:F(lambda)}
%   F(\lambda)=\int_0^\infty \frac{\cosh(\lambda x)x^4dx}
%                               {\cosh x-\cosh(\lambda x)},
% \quad
%   -1<\lambda<1.
% \end{equation}
% Remarkably, $\chi(q)$ is fully determined by the phonon spectrum and its
% dependence on particle density $n$.
% In the most important case of $q=\pi$ we have $v_q=0$, and the expression
% (\ref{eq:chi(q)}) takes the form
\begin{eqnarray}
  \label{eq:chi}
  \chi=\frac{4\pi^3
            [(\partial_n\omega_\pi)\partial_nv^2
            -v^2\partial_n^2\omega_\pi
            +\omega_\pi'' (\partial_n\omega_\pi)^2n^{-2}]^2
            }{15\hbar^3m^2v^{12}},
\nonumber\\
\end{eqnarray}
Remarkably, $\chi$ is fully determined by the phonon spectrum and its
dependence on the particle density $n$.
% where we used $F(0)=8\pi^4/15$.

Equations (\ref{eq:relaxation_rate}), (\ref{eq:B_vs_T}), and (\ref{eq:chi})
give the complete expression for the equilibration rate of a one-dimensional
Wigner crystal.  In the case of pure Coulomb repulsion $V(x)=e^2/|x|$ the
velocity $s$ diverges, and our treatment is inapplicable.  However, in the
experimental realizations of one-dimensional Wigner crystal, there is usually
a metal gate screening the interactions at large distances.  In this case we
obtain
\begin{equation}
  \label{eq:rate_Coulomb}
  \frac1\tau=\eta
             \frac{\Delta}{\hbar\ln^{5/2}(d/a)}
             \frac{a_B}{a}
             \left(\frac{T}{\Delta}\right)^{3/2}
             e^{-\Delta/T},
\end{equation}
where $\eta=63\pi^3\zeta(3)\sqrt{\ln2}/80\sqrt{2\pi}$ is a numerical
prefactor, $a_B=\hbar^2/me^2$ is the Bohr's radius, $d$ is the distance to the
gate, and
\begin{equation}
  \label{eq:Delta}
  \Delta=\hbar\omega_\pi=\left(\frac{7\zeta(3)\hbar^2e^2}{ma^3}\right)^{1/2}.
\end{equation}

A non-trivial test of our result (\ref{eq:chi}) can be performed by
considering the interaction potential 
\begin{equation}
  \label{eq:sinh_interaction}
  V(x)=\frac{\gamma}{\sinh^2 cx}.
\end{equation}
It is well known \cite{sutherland} that the model (\ref{eq:sinh_interaction})
is integrable, i.e., it has an infinite number of integrals of motion.  As a
result the excitations of the system have infinite lifetimes, and one expects
the diffusion constant $B$ to vanish.  Our approach applies only to the limit
of strong repulsion $\gamma\to\infty$, when the Wigner crystal approximation
is applicable.  On the other hand, the parameter $c$ can take any value.  It
is easy to obtain analytic expressions for the phonon spectrum in the limiting
cases $c\ll n$ and $c\gg n$.

At $c\ll n$ one can approximate (\ref{eq:sinh_interaction}) with
$V(x)=\gamma/(cx)^2$, the so-called Calogero-Sutherland model
\cite{sutherland}.  Then from Eq.~(\ref{eq:omega_q}) one finds
\begin{eqnarray}
  \label{eq:omega_calogero}
  \omega_q&=&\left(\frac\gamma{c^2m}\right)^{1/2}
             n^2 (\pi q-q^2/2),
\\
  \label{eq:v_q_calogero}
  v_q&=&\left(\frac\gamma{c^2m}\right)^{1/2}
        n(\pi-q),
\end{eqnarray}
and $v$ is given by $v_q$ at $q=0$.  Substitution of
Eqs.~(\ref{eq:omega_calogero}) and (\ref{eq:v_q_calogero}) into
(\ref{eq:Upsilon}) gives $\Upsilon_q=0$, and therefore $B=0$.

At $c\gg n$ the interactions fall off very rapidly with the distance,
$V(x)=4\gamma e^{-2cx}$.  In this case only the interaction of the nearest
neighbor particles in the Wigner crystal needs to be taken into account (Toda
lattice).  The spectrum takes the form
\begin{eqnarray}
  \label{eq:omega_toda}
  \omega_q&=&8c\left(\frac\gamma{m}\right)^{1/2}
           e^{-c/n} \sin\frac{q}{2},
\\
  \label{eq:v_q_toda}
  v_q&=&4c\left(\frac\gamma{m}\right)^{1/2}
           \frac{1}{n}\,e^{-c/n} \cos\frac{q}{2}.
\end{eqnarray}
As expected, substitution of Eqs.~(\ref{eq:omega_toda}) and
(\ref{eq:v_q_toda}) into (\ref{eq:Upsilon}) gives $\Upsilon_q=0$.  Finally, we
have checked numerically that the expression (\ref{eq:chi}) vanishes for
potential (\ref{eq:sinh_interaction}) for any $c$.

To summarize, we have obtained the equilibration rate $\tau^{-1}$ of
one-dimensional system of particles with strong long-rage repulsion.  At low
temperatures the rate is exponentially suppressed with the activation energy
given by the Debye frequency $\omega_\pi$ of the Wigner crystal.  The
prefactor can be expressed in terms of the phonon spectrum using
Eqs. (\ref{eq:relaxation_rate}), (\ref{eq:B_vs_T}), and (\ref{eq:chi}).  In
the case of Coulomb repulsion the result is given by
Eqs.~(\ref{eq:rate_Coulomb}) and (\ref{eq:Delta}).  Finally, we have checked
that the equilibration rate vanishes for the integrable model of particles
with interactions in the form (\ref{eq:sinh_interaction}).

The authors are grateful to A. Levchenko for helpful comments.  This work was
supported by the U.S. Department of Energy under Contracts
No. DE-AC02-06CH11357 and DE-FG02-07ER46452.

\end{document}